\documentclass[letter]{aa} 

\usepackage{graphicx}
\usepackage{txfonts}
\begin{document}

   \title{Candidate exoplanet host HD\,131399A: a nascent Am star}

   \author{N. Przybilla\inst{1}
          \and
          P. Aschenbrenner\inst{1}
          \and 
          S. Buder\inst{2}
          }

   \institute{Institut f\"ur Astro- und Teilchenphysik, Universit\"at Innsbruck,
              Technikerstrasse 25, 6020 Innsbruck, Austria\\
              \email{norbert.przybilla@uibk.ac.at}  
         \and
              Max-Planck-Institut f\"ur Astronomie,
              K\"onigstuhl 17, 69117 Heidelberg, Germany
             }

   \date{Received ; accepted }

 
  \abstract
   {Direct imaging suggests that there is a Jovian exoplanet 
   around the primary A-star in
   the triple-star system HD\,131399. 
   We investigate a high-quality spectrum 
    of the primary component HD\,131399A obtained with FEROS on the
    ESO/MPG 2.2m telescope, aiming to
   characterise the star's atmospheric and fundamental parameters, and to
   determine elemental abundances at high precision and accuracy. 
   The aim is to constrain the chemical composition of the birth cloud 
   of the system and therefore the bulk composition of the putative planet.
   A hybrid non-local thermal equilibrium (non-LTE) model atmosphere technique is adopted for the
   quantitative spectral analysis.
   Comparison with the most recent stellar evolution models yields 
   the fundamental parameters. 
   The atmospheric and fundamental stellar parameters of HD\,131399A are 
   constrained to $T_\mathrm{eff}$\,=\,9200$\pm$100\,K, $\log
   g$\,= 4.37$\pm$0.10, $M$\,=\,1.95$^{+0.08}_{-0.06}$\,$M_\odot$,
   $R$\,=\,1.51$^{+0.13}_{-0.10}$\,$R_\odot$ , and $\log L/L_\odot$,=\,1.17$\pm$0.07,
   locating the star on the zero-age main sequence.
   Non-LTE effects on the derived metal abundances are often
   smaller than 0.1\,dex, but can reach up to $\sim$0.8\,dex for
   individual lines. The observed lighter elements up to calcium are overall
   consistent with present-day cosmic abundances, 
   with a C/O ratio of 0.45$\pm$0.07 by number, while  
   the heavier elements show mild overabundances.
   We conclude that the
   birth cloud of the system had a standard chemical composition, but
   we witness the onset of the Am phenomenon in the slowly rotating star.
   We furthermore show that non-LTE analyses have the potential to solve
   the remaining discrepancies between observed abundances and predictions by
   diffusion models for Am stars. Moreover, the present case
   allows mass loss, not turbulent mixing, to be identified as the main
   transport process competing with diffusion in very young Am stars.} 

   \authorrunning{Przybilla et al.}

   \maketitle
%

\section{Introduction}
The source HD\,131399 is a member of the 16$\pm$7\,Myr old Upper Centaurus-Lupus association
\citep[UCL,][]{deZeeuwetal99,PeMa16}. The hierarchical triple system was only
recently resolved using SPHERE on the ESO VLT by
\citet{Wagneretal16}. It consists of the A-type primary HD\,131399A and a
$\sim$3.2\arcsec~distant close pair HD\,131399BC, photometrically
classified as a G and a K dwarf.
Moreover, the study suggested that a Jovian exoplanet,
HD\,131399Ab, is located at $\sim$0.8\arcsec~distance from the primary. This is only the 
fourth and hottest A-type exoplanet host system detected by 
direct imaging \citep[note that an alternative interpretation of HD\,131399Ab as a
background star has been proposed in the meantime by][]{Nielsenetal17}.

Little is known about the stars of the system, with no dedicated study
to be found in the literature. Estimates of atmospheric and fundamental parameters 
of the A-star primary were mostly published as part of photometry-based
determinations of the properties of large star samples and one orbital
study of the system, see Table~\ref{tab:lit} for a summary. Evidently,
the scatter in the parameter values is quite large. No high-resolution spectroscopy
has been obtained so far to determine elemental abundances.

The close proximity of the HD\,131399BC components of
$<$0.1\arcsec~complicates a quantitative study of the
individual stars. To constrain the chemical composition of
the birth cloud of the system, for example, as a tracer for the bulk composition of
the suggested planet, the investigation therefore has to concentrate on
HD\,131399A. This was the original motivation for the present
work, but it developed into an unexpected direction.

\begin{table}[t]
\caption[]{Literature data for the parameters of HD\,131399A.\\[-9.5mm]}
\label{tab:lit}
\setlength{\tabcolsep}{0.8mm}
\begin{center}
\begin{footnotesize}
\begin{tabular}{ll}
\hline\hline
\citet{deGeusetal89}  & $T_\mathrm{eff}$\,=\,8910\,K, $\log g$\,(cgs)\,=\,4.38,\\
                      & $\log L/L_\odot$\,=\,1.2, $A_V$\,=\,0.14\,mag\\[.5mm]
\citet{Kouwenhovenetal07}& $M$\,=\,1.82\,$M_\odot$\\[1mm]
\citet{McDonaldetal12}& $T_\mathrm{eff}$\,=\,7946\,K, $\log L/L_\odot$\,=\,1.064\\[.5mm]
\citet{DaHi15}        & $T_\mathrm{eff}$\,=\,9324$\pm$317\,K, $\log g$\,=\,4.48\\[.5mm] 
\citet{Nielsenetal17} & $T_\mathrm{eff}$\,=\,9480$^{+420}_{-410}$\,K, $\log g$\,=\,4.32$\pm$0.01,\\
                      & $A_V$\,=\,0.22$\pm$0.09\,mag,
                      $M$\,=\,2.08$^{+0.12}_{-0.11}$\,$M_\odot$\\[.5mm]
\hline\\[-5mm]
\end{tabular}
\end{footnotesize}
\end{center}
\end{table}


\begin{figure*}
\includegraphics[width=.995\linewidth]{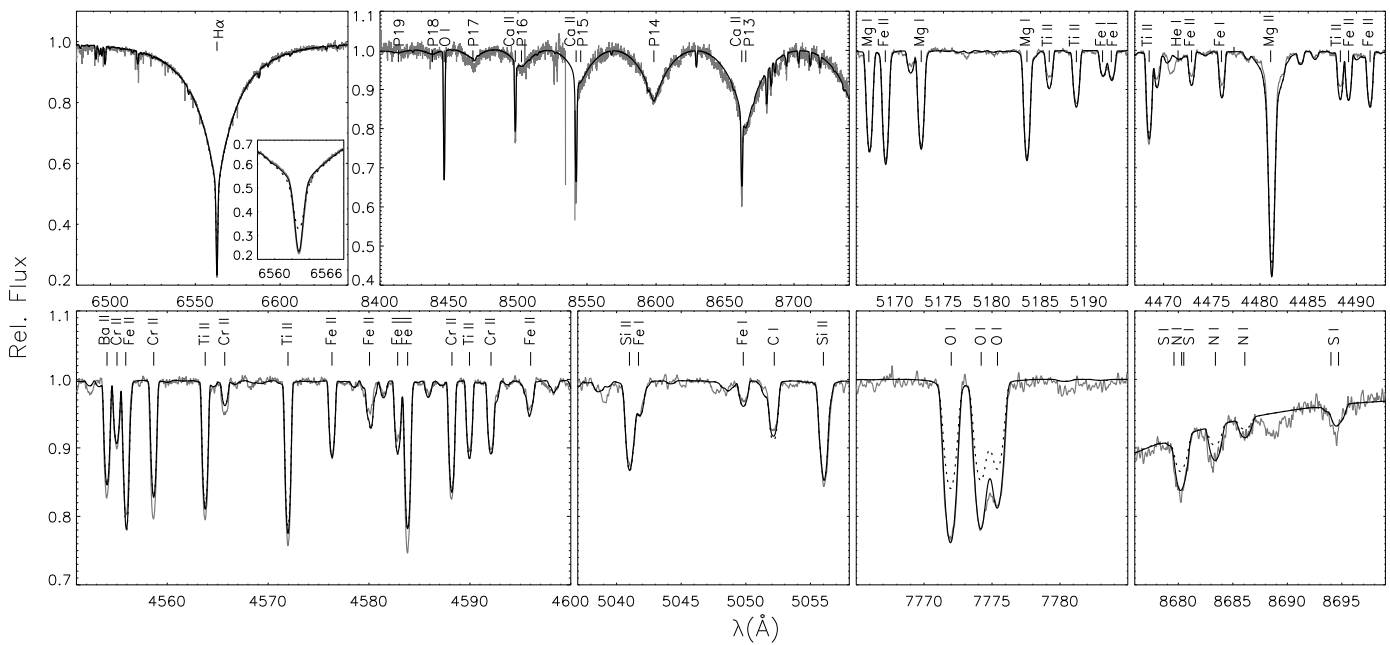}
\caption{Comparison of the spectrum of HD\,131399 (grey) 
with the model computed for the final atmospheric
parameters and abundances (full line) for several diagnostic
regions. Pure LTE profiles (dotted) are also indicated to visualise
the non-LTE effects. The inset shows the H$\alpha$ core.\label{fig:fits}}
\end{figure*}

\section{Observations and data reduction}
We observed HD\,131399A with the Fiber-fed Extended Range Optical Spectrograph (FEROS)
on the ESO/MPG 2.2m telescope at La Silla on March 12, 2017. The measured DIMM seeing during
observation was $\sim$0.6{\arcsec}, resulting in no significant light
from the $\sim$3.2\arcsec~distant HD\,131399BC pair falling into 
the 2\arcsec~fibre aperture. Near-complete wavelength coverage 
from 3700 to 9200\,{\AA} was achieved at resolving power 
$R$\,=\,$\lambda/\Delta\lambda$\,= 48\,000, with a signal-to-noise
ratio $S/N$\,$\approx$\,250 at 
$\lambda$\,=\,5000\,{\AA} in the 600\,s exposure. 

The basic data reduction was performed using the FEROS data reduction system
\citep[for details see][]{Kauferetal99}. The spectrum was normalised by
fitting a spline function to carefully selected continuum points,
taking into account that the Balmer lines are near maximum strength in
HD\,131399A. Finally, the spectra were shifted to the laboratory rest
frame by accounting for the measured radial velocity $v_\mathrm{rad}$
(see Table~\ref{tab:parameters})
as determined from cross-correlation with an appropriate synthetic spectrum.


\section{Model atmosphere analysis}
We employed a hybrid non-LTE approach for the model atmosphere 
analysis of HD\,131399A as introduced to A-type stars by
\citet{Przybillaetal06} and used later also for the analysis of
chemically peculiar early-type stars \citep{Przybillaetal08,Przybillaetal16}. 
This is based on plane-parallel
homogeneous model atmospheres in hydrostatic, radiative, and 
local thermodynamic equilibrium (LTE) as computed with the 
{\sc Atlas9} code \citep{Kurucz93}. In a subsequent step, non-LTE
line-formation calculations were performed with recent versions of {\sc Detail}
and {\sc Surface} \citep[both updated and extended by K. Butler]{Giddings81,BuGi85}.
The coupled radiative transfer and statistical
equilibrium equations were solved with {\sc Detail}, employing
the accelerated lambda iteration scheme of \citet{RyHu91}. 
{\sc Surface} was then used to calculate non-LTE and LTE synthetic spectra based on
refined line-broadening theories.
Continuous opacities due to hydrogen and helium were considered in non-LTE, and
line blocking was accounted for via LTE opacity sampling, employing
the method of \citet{Kurucz96}. 

Lines for a range of important chemical species (H, He, C, N, O, Mg, Si, Ti, and Fe) 
were treated in non-LTE (and LTE for quantifying 
the non-LTE--LTE abundance differences). Model atoms as discussed by
\citet{Przybillaetal06,Przybillaetal16} were adopted. Spectral lines 
of 16 additional chemical elements were treated in LTE.

We employed {\sc Spas} \citep[Spectrum Plotting and Analysing
Suite,][]{Hirsch09} for the comparison of synthetic spectra with
observation. {\sc Spas} allows
interpolating between model grid points for up to three parameters 
simultaneously, and instrumental and rotational
broadening can be applied to the theoretical
profiles. The program uses a downhill simplex algorithm to minimise $\chi^2$. 

The atmospheric parameter determination was based on simultaneous 
fits to the Stark-broadened hydrogen lines and the \ion{Mg}{i/ii}
ionisation equilibrium (requiring that the two ionisation stages
indicate the same abundance) in order to constrain effective temperature
$T_\mathrm{eff}$ and surface gravity $\log g$
\citep[see e.g.][]{Przybillaetal06}. The microturbulence velocity $\xi$ was adjusted
such that the abundances of the species treated in non-LTE became independent 
of the strength of the spectral lines. Line-profile fits were used for
both the determination of the elemental abundances and of the 
projected rotational velocity $v_\mathrm{rot} \sin i$. 
The results of the model atmosphere analysis are summarised
in Table~\ref{tab:parameters}. Elemental abundances are 
averages over all analysed lines of a chemical species, giving equal
weight to each spectral feature, except for manganese and iron, where 
only the results for the ions are considered. Uncertainties are 
1$\sigma$ standard deviations from the line-to-line scatter.
Individual line abundances in non-LTE and/or LTE plus more detailed information on ionic
abundances can be found in the Appendix.

Based on the final atmospheric parameters and abundances, a synthetic spectrum
was calculated that is compared to a selection of observed diagnostic
features in Figure~\ref{fig:fits}. A similar match is achieved for
practically the entire observed spectrum, the deviations reflect the
line-to-line scatter of abundances. 
Table~\ref{tab:parameters} also provides some general information on
HD\,131399A, such as the spectral type and the 
Hipparcos parallax, distance, and proper-motion data. We note that no data
on HD\,131399 were provided by the first Gaia data release
\citep{Gaia16} as a consequence of its multiplicity.

Starting from the photometric measurements of \citet[we note that the
contribution of HD\,131399BC is insignificant in the optical]{Slawsonetal92}
we furthermore constrained the interstellar reddening $E(B-V)$ and extinction $A_V$ 
(adopting a standard ratio of total-to-selective extinction of 3.1) by
comparison with synthetic photometry of the model flux. From these and the known
distance, we determined the absolute visual magnitude $M_V$ of HD\,131399A,
and by adopting the bolometric correction from our model atmosphere,
the absolute bolometric magnitude $M_\mathrm{bol}$ (see
Table~\ref{tab:parameters}). The uncertainties
of both quantities are entirely dominated by the error in the
Hipparcos parallax/distance.

\begin{table}[t]
\caption[ ]{Parameters and elemental abundances of HD\,131399A.\\[-9mm]}
\label{tab:parameters}
\setlength{\tabcolsep}{0.5mm}
\begin{center}
\begin{footnotesize}
\begin{tabular}{ll@{\hspace{1.7mm}}ll}
\hline\hline
\multicolumn{4}{l}{General information}\\
Sp.~Type          & A1\,V\\
$v_\mathrm{rad}$  & 0.51$\pm$0.10\,km\,s$^{-1}$ & $\pi$        & 10.20$\pm$0.70\,mas $[$1$]$\\                     
$d_\mathrm{HIP}$  & 98.0$^{+7.3}_{-6.3}$\,pc    & $\mu_\alpha$ & $-$29.69$\pm$0.59\,mas\,yr$^{-1}$ $[$1$]$\\
$d_\mathrm{spec}$ & 98.0$\pm$13.0\,pc           & $\mu_\delta$ & $-$31.52$\pm$0.55\,mas\,yr$^{-1}$ $[$1$]$\\[1.2mm]
\multicolumn{4}{l}{Atmospheric parameters}\\
$T_\mathrm{eff}$           & 9200$\pm$100\,K & $\xi$      & 2$\pm$0.5\,km\,s$^{-1}$\\
$\log g$\,(cgs)            & 4.37$\pm$0.10   & $v_\mathrm{rot} \sin i$ & 26$\pm$2\,km\,s$^{-1}$\\[1.2mm]
\multicolumn{4}{l}{Elemental abundances $\log$\,($X$/H)\,+\,12}\\
He     & 10.99\,(1)          &  V$^*$  &  4.00$\pm$0.17\,(2) \\ 
C      & 8.38$\pm$0.06\,(15) &  Cr$^*$ &  5.94$\pm$0.11\,(30)\\ 
N      & 7.65$\pm$0.05\,(6)  &  Mn$^*$ &  5.88.$\pm$0.10\,(2)\\       
O      & 8.73$\pm$0.04\,(7)  &  Fe     &  7.70.$\pm$0.10\,(34)\\      
Na$^*$ & 6.30$\pm$0.03\,(3)  &  Co$^*$ &  5.32$\pm$0.03\,(2)\\       
Mg     & 7.55$\pm$0.05\,(13) &  Ni$^*$ &  6.53$\pm$0.10\,(7)\\       
Al$^*$ & 6.19$\pm$0.07\,(2)  &  Cu$^*$ &  4.47$\pm$0.11\,(2)\\       
Si     & 7.57$\pm$0.04\,(7)  &  Zn$^*$ &  5.16$\pm$0.06\,(2)\\             
S$^*$  & 7.22$\pm$0.14\,(4)  &  Sr$^*$ &  3.18$\pm$0.03\,(2)\\      
Ca$^*$ & 6.23$\pm$0.14\,(20) &  Y$^*$  &  2.14$\pm$0.23\,(3)\\
Sc$^*$ & 3.20$\pm$0.10\,(5)  &  Zr$^*$ &  3.05$\pm$0.08\,(3)\\
Ti     & 5.51$\pm$0.17\,(29) &  Ba$^*$ &  3.00$\pm$0.18\,(5)\\[1.2mm]
\multicolumn{3}{l}{Photometric data}\\
$V$   & 7.048$\pm$0.008\,mag $[$2$]$     & $A_V$            & 0.202$\pm$0.012\,mag\\
$B-V$ & 0.118$\pm$0.007\,mag $[$2$]$     & $M_V$            & 1.89$^{+0.37}_{-0.32}$\,mag\\
$E(B-V)$         & 0.065$\pm$0.012\,mag  & $M_\mathrm{bol}$ & 1.76$^{+0.37}_{-0.32}$\,mag\\[1.2mm]
\multicolumn{3}{l}{Fundamental parameters}\\
$M/M_\odot$                & 1.95$^{+0.08}_{-0.06}$ & $\log L/L_\odot$ & 1.17$\pm$0.07\\
$R/R_\odot$                & 1.51$^{+0.13}_{-0.10}$ & $\tau$           & 16$\pm$7\,Myr $[$3$]$\\
\hline\\[-5mm]
\end{tabular}
\tablefoot{1$\sigma$ uncertainties are given. For abundances these are
from the line-to-line scatter. Numbers in brackets quantify the
analysed lines. LTE abundance values are marked by an
asterisk. References are indicated by square brackets.}
\tablebib{(1)~\citet{vanLeeuwen07}; (2)~\citet{Slawsonetal92};
(3)~\citet{PeMa16}.}
\end{footnotesize}
\end{center}
\end{table}

\begin{figure}
\centering
\includegraphics[width=0.90\linewidth]{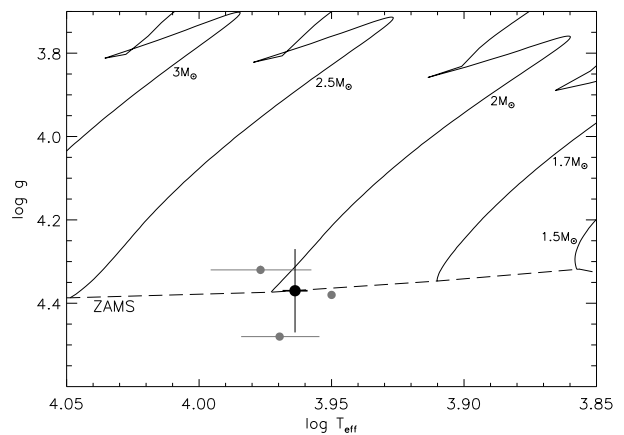}
\caption{HD\,131399A in the $\log T_\mathrm{eff}$--$\log g$ diagram
(black dot). Overlaid are Geneva evolution tracks for non-rotating
stars, computed for metallicity $Z$\,=\,0.014 
\citep[full lines]{Ekstroemetal12}, spanning the width of 
the main sequence. The dashed line marks the zero-age main sequence. 
Results from previous parameter determinations (Table~\ref{tab:lit}) 
are marked by grey dots. Where available, 1$\sigma$ error bars are 
displayed.   \label{fig:evol}}
\end{figure}

In order to constrain the fundamental parameters of HD\,131399A, we
employed the stellar evolution models of \citet{Ekstroemetal12}. The
tracks for non-rotating stars were chosen, motivated by the finding
that HD\,131399A is a slowly rotating star, as discussed in
the next section. Figure~\ref{fig:evol} shows the location of
HD\,131399A in the $\log T_\mathrm{eff}$--$\log g$ diagram, which covers
the whole evolutionary sequence from the zero-age main sequence (ZAMS) to the end of
core hydrogen burning. This facilitated the (evolutionary) mass $M$ to be
determined. Based on this, the stellar radius $R$ and luminosity $L$
were constrained using the derived atmospheric parameters, see
Table~\ref{tab:parameters}. In addition, the spectroscopic distance
$d_\mathrm{spec}$ was derived employing Eq.~1 of \citet{NiPr12}.
Its uncertainty is dominated by the error bars of $\log g$.
The spectroscopic distance is in excellent agreement with the 
independent Hipparcos measurement, thus ensuring that a consistent 
solution for the atmospheric and fundamental parameters was found. 

Because of the position of HD\,131399A on the ZAMS -- in agreement with expectation --
we cannot employ an isochrone fit to determine its age $\tau$.
Instead, the independent age estimate for the UCL
association by \citet{PeMa16} was adopted, which accounts for the
intrinsic age spread of the members of UCL. From this and the stellar
evolution models, we find that HD\,131399A has spent only $\sim$1\% of its
lifetime so far.


\section{Discussion}
Our values from high-resolution spectroscopy are compatible
with previous photometry-based atmospheric parameter estimates
(Fig.~\ref{fig:evol}), but with significantly reduced uncertainties in
$T_\mathrm{eff}$. Our observations facilitate a determination of  
$v_\mathrm{rot} \sin i$ for the first time. 
Even though \citet{Wagneretal16} provided only loose constraints on the
inclination angle from their orbital characterization of the system,
$i$\,=\,45\degr~to 65\degr~for the orbit of HD\,131399BC
around HD\,131399A, this is sufficient to place an upper
limit of $v_\mathrm{rot}$\,$<$\,40\,km\,s$^{-1}$ assuming a
co-alignment of the star's rotational axis with the axis of the
orbital motion. This means that HD\,131399A is a slowly rotating star.

\begin{figure*}
\sidecaption
\includegraphics[width=12cm,height=4.7cm]{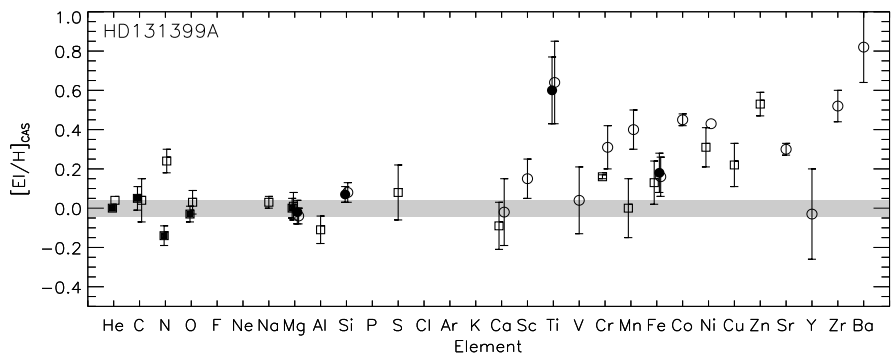}
\caption{Elemental abundances in HD\,131399A, relative to the cosmic
abundance standard \citep[CAS][]{NiPr12,Przybillaetal13}, supplemented
by solar meteoritic abundances \citep{Asplundetal09} where unavailable
from young massive stars. The grey bar spans the expected CAS range. 
Non-LTE abundances for HD\,131399A are
indicated by filled, and LTE values by open symbols. Squares mark data based
on neutral species, and circles data from single-ionized species. 
Error bars are 1$\sigma$ uncertainties as 
deduced from from the line-to-line scatter.\label{fig:abus}}
\end{figure*}

We present the most complete non-LTE abundance determination in an A star
to date. A third of the observed chemical species and about half of the total
spectral lines was treated in non-LTE, thus improving the accuracy and
precision of the resulting abundance data.

As a very young and nearby star, we would expect HD\,131399 to show pristine
abundances such as those typical today in the solar
neighbourhood \citep{NiPr12,Przybillaetal13}. Figure~\ref{fig:abus} shows a comparison
of the derived abundances with this cosmic abundance standard (CAS), 
supplemented by meteoritic solar abundances
\citep{Asplundetal09} where data from young massive stars are unavailable. 
The expected range of abundance values is highlighted by the grey
band, with some deviations to be considered for elements where only LTE values
are available. The lighter elements up to Ca are overall
compatible with expectation. In particular, the C/O ratio is found to be
0.45$\pm$0.07 by number, in good agreement with the CAS average of
0.43$\pm$0.06. However, marked overabundances become apparent for
the heavier elements, up to a factor of $\sim$6 for Ba.
Enrichment levels for the lanthanides are apparently not high enough
to produce detectable lines. 

As an early A-type star, HD\,131399A has a very thin surface
convection zone that is prone to develop chemical peculiarities for
$v_\mathrm{rot}$\,$\lesssim$\,120\,km\,s$^{-1}$ \citep{Abt00}. This
can be a fast process occurring on timescales as short as some 10$^6$
years, that is, even in the pre-main-sequence-phase \citep{Vicketal11}. 
The reason is that atomic diffusion \citep[e.g.][]{Michaud70} 
is apparently important in HD\,131399A even though it has avoided
classification as an Am star.
HD\,131399A does not show the abundance pattern characteristic
for Am stars -- marked underabundances for C and Sc, underabundances
for O and Ca, and overabundances of the iron-group elements -- as
discussed by \citet[and references therein]{Vicketal10}, for
instance.
However, typical Am
stars have ages of the order of several 10$^8$\,yr, while the spectral
characteristics are still developing in HD\,131399A.

Stellar models accounting for atomic diffusion may provide some
guidance for the interpretation. Figure~15 of
\citet{Richeretal00} and Fig.~14 of \citet{Vicketal10}, for example, discuss the
evolution of the surface abundances as a function of age. While these
models are not tailored to the parameters of HD\,131399A,
they should be close enough to describe the effects qualitatively.
The models of Richer et al. and Vick et al. 
in the age range of HD\,131399A indeed show the observed 
pattern, nearly pristine abundances up to calcium, and mild
overabundances of the iron-group elements. Models with either
turbulent mixing or mass loss as the main competitor to atomic
diffusion make very similar predictions for Am stars at later ages
\citep{Michaudetal11}, but there are some distinct features at early times.
In particular, a large overabundance of titanium
is unique \citep{Vicketal10}. We can therefore identify
mass loss as the main transport process counter-acting diffusion
in very young Am stars. We conclude that HD\,131399A can be viewed 
as a rare but very valuable case of a {\em \textup{nascent Am star}}. The present work provides
tight observational constraints for atomic diffusion in very
young stars that start to develop the Am phenomenon. We would like to suggest further
investigations of the fast development of the titanium abundance in
diffusion models, as this could provide an accurate and precise age
indicator for nascent Am stars, a `titanium clock', and
a means to constrain otherwise unaccessible mass-loss rates in young A-type stars. 

Concerning our initial motivation for this study, we may conclude 
that it is probable that the pristine abundances for the heavier elements of the
iron group and beyond were also compatible with CAS/solar abundances
in view of the ongoing diffusion processes.
The birth cloud out of which the entire system HD\,131399 including the
candidate planet formed  most likely showed a standard composition
as typical for the present-day solar neighbourhood. 

Finally, we would like to discuss the non-LTE effects on the abundance
analysis, as these were studied to such large extent in an Am star 
for the first time. This is visualised in Fig.~\ref{fig:fits}, where
the differences between non-LTE and LTE model spectra are shown, in
Fig.~\ref{fig:abus}, where the ionic abundances in non-LTE and LTE are compared, 
and in Table~\ref{tab:lines} in the Appendix, where the
details from the line-by-line analysis are summarised.
Non-LTE effects on average abundances are
usually small in HD\,131399 for most of the elements. They amount to
$\lesssim$0.1\,dex, {\em if} some spectral lines with obviously
discrepant LTE abundance values are not considered. These outliers include
classical features such as the \ion{O}{i} $\lambda\lambda$\,7771-5\,{\AA} triplet
and \ion{O}{i} $\lambda$\,8446\,{\AA}, which show a pronounced 
non-LTE--LTE abundance difference $\Delta\log\varepsilon$ by $\sim$0.8\,dex 
\citep[for a detailed discussion see e.g.][]{Przybillaetal00}.
Other (less extreme) examples are the red
\ion{Si}{ii} $\lambda\lambda$\,6347,6371\,{\AA} lines, which show
$\Delta\log\varepsilon$\,$\approx$\,0.25\,dex. 
However, for \ion{N}{i,} all lines are affected almost uniformly by 
rather large non-LTE effects, $\Delta\log\varepsilon$\,$\approx$\,0.4\,dex
\citep[see e.g.][]{PrBu01}. This has the
potential to solve one of the most prominent discrepancies between observation 
and predictions from diffusion models for Am stars \citep[and
references therein]{Richeretal00,Vicketal10}.
Discrepancies in the ionisation equilibria of some elements that can
be currently analysed in LTE only hint on the presence of (substantial) non-LTE
effects for these as well.

We have to state, however, that the differences between non-LTE and LTE
abundances found here should not be generalised throughout the Am star parameter
space, but trends can be estimated. Non-LTE effects might well become
stronger for the typically (much) more evolved Am stars. On the one hand,
the lower-density plasma at lower gravities supports less thermalising
collisions, but on the other hand, the overall higher atmospheric
abundances for most elements will extend the line-formation region
outwards in the atmospheres, where non-LTE departures are more
pronounced. We conclude that additional non-LTE studies of Am stars
(also covering more elements) will
be worthwhile to provide unbiased abundances for the comparison with
diffusion models, which may in turn open up possibilities for 
refining the models as well.


\begin{acknowledgements}
SB acknowledges funding from the Alexander von Humboldt Foundation
in the framework of a Sofja Kovalevskaja Award endowed by
the German Federal Ministry of Education and Research.
\end{acknowledgements}

%
%

\begin{appendix}
\section{Line-by-line analysis}
Table~\ref{tab:lines} summarises the results from the abundance analysis of
individual lines of the different chemical species observed in
HD\,131399A. For each ion the transition wavelengths $\lambda$ are given, 
together with the excitation potential of the lower level $\chi$, 
the adopted oscillator strength $\log gf$ for the transition, an accuracy indicator 
and the reference to the source of the oscillator strength, the
abundance $\log \varepsilon$\,=\,$\log$\,($X$/H)\,+\,12 (in non-LTE,
if a model atom is available, otherwise in LTE), and the difference
between non-LTE and LTE abundance, 
$\Delta\log \varepsilon$\,=\,$\log \varepsilon_\mathrm{NLTE}- \log \varepsilon_\mathrm{LTE}$.
When several fine-structure transitions contribute to an
observed line, abundance data are given at the first entry only.
For each ion, the total abundance is also indicated in non-LTE and/or
LTE, averaged using the same weight for all lines, and the
1$\sigma$ uncertainty from the line-to-line scatter. The number of
analysed lines is given in brackets. Some lines show obvious non-LTE
effects when compared to the other transitions in an ion. In these
cases, the line is marked in italics and is discarded from the LTE 
average, as would be done in a classical LTE study.

\begin{table}[bh!]
\caption[]{Line-by-line elemental abundances\\[-9mm]}
\label{tab:lines}
\setlength{\tabcolsep}{2mm}
\begin{center}
\begin{footnotesize}
\begin{tabular}{lrrlrrr}
\hline\hline
$\lambda$\,({\AA}) & $\chi$\,eV) & $\log gf$ & Acc. & Ref. & $\log \varepsilon$ & $\Delta \log \varepsilon$\\
\hline
\multicolumn{7}{l}{\ion{He}{i}:\hfill$\log
\varepsilon_\mathrm{NLTE}$\,=\,10.99\,(1),~~~
$\log \varepsilon_\mathrm{LTE}$\,=\,11.03\,(1)}\\[0.5mm]
 5875.599 & 20.96 & $-$1.516 & AAA  & WF   &10.99 & $-$0.04\\
 5875.614 & 20.96 & $-$0.340 & AAA  & WF \\
 5875.615 & 20.96 &    0.409 & AAA  & WF \\
 5875.625 & 20.96 & $-$0.339 & AAA  & WF \\
 5875.640 & 20.96 &    0.138 & AAA  & WF \\
 5875.966 & 20.96 & $-$0.214 & AAA  & WF \\[1mm]
\multicolumn{7}{l}{\ion{C}{i}:\hfill$\log
\varepsilon_\mathrm{NLTE}$\,=\,8.38$\pm$0.06\,(15),~~~
$\log \varepsilon_\mathrm{LTE}$\,=\,8.37$\pm$0.11\,(15)}\\[0.5mm]
 4771.742 &  7.49 & $-$1.866 & C    & WFD  & 8.45 &    0.02\\
 4775.898 &  7.49 & $-$2.304 & C    & WFD  & 8.33 &    0.02\\
 4932.049 &  7.68 & $-$1.658 & B    & WFD  & 8.37 &    0.06\\
 5052.167 &  7.68 & $-$1.303 & B    & WFD  & 8.31 &    0.07\\
 5668.951 &  8.54 & $-$1.429 & C    & LP   & 8.37 &    0.03\\
 6013.166 &  8.65 & $-$1.314 & D    & WFD  & 8.37 &    0.05\\
 6013.213 &  8.65 & $-$1.673 & D    & WFD  \\
 6014.830 &  8.64 & $-$1.585 & D    & WFD  & 8.34 &    0.02\\
 6587.610 &  8.54 & $-$1.003 & B    & WFD  & 8.32 &    0.09\\
 7100.124 &  8.64 & $-$1.470 & B    & WFD  & 8.47 &    0.03 \\
 7108.934 &  8.64 & $-$1.592 & B    & WFD  & 8.38 &    0.02\\
 7111.472 &  8.64 & $-$1.086 & B$-$ & WFD  & 8.49 &    0.02\\
 7113.178 &  8.65 & $-$0.774 & B$-$ & WFD  & 8.29 &    0.05\\
 9078.288 &  7.48 & $-$0.581 & B    & WFD  & 8.40 & $-$0.09\\
 9088.513 &  7.48 & $-$0.429 & B    & WFD  & 8.39 & $-$0.11\\
 9111.807 &  7.49 & $-$0.298 & B    & WFD  & 8.44 & $-$0.14\\[1mm]
\multicolumn{7}{l}{\ion{N}{i}:\hfill$\log
\varepsilon_\mathrm{NLTE}$\,=\,7.65$\pm$0.05\,(6),~~~
$\log \varepsilon_\mathrm{LTE}$\,=\,8.03$\pm$0.06\,(6)}\\[0.5mm]
 7442.298 & 10.33 & $-$0.384 & B+   & WFD  & 7.59 & $-$0.45\\
 8594.000 & 10.68 & $-$0.334 & B    & WFD  & 7.59 & $-$0.38\\
 8629.235 & 10.69 &    0.075 & B    & WFD  & 7.61 & $-$0.42 \\
 8683.403 & 10.33 &    0.086 & B+   & WFD  & 7.70 & $-$0.44 \\
 8686.149 & 10.33 & $-$0.305 & B+   & WFD  & 7.63 & $-$0.35 \\
 8703.247 & 10.33 & $-$0.322 & B+   & WFD  & 7.69 & $-$0.34\\[1mm]
\multicolumn{7}{l}{\ion{O}{i}:\hfill$\log
\varepsilon_\mathrm{NLTE}$\,=\,8.73$\pm$0.04\,(7),~~~
$\log \varepsilon_\mathrm{LTE}$\,=\,8.79$\pm$0.06\,(4)}\\[0.5mm]
 6155.961 & 10.74 & $-$1.363 & B+   & WFD  & 8.67 & $-$0.05\\
 6155.971 & 10.74 & $-$1.011 & B+   & WFD \\
 6155.989 & 10.74 & $-$1.120 & B+   & WFD \\
 6156.737 & 10.74 & $-$1.487 & B+   & WFD \\
 6156.755 & 10.74 & $-$0.898 & B+   & WFD \\
 6156.788 & 10.74 & $-$0.694 & B+   & WFD \\
 6158.149 & 10.74 & $-$1.841 & B+   & WFD  & 8.72 & $-$0.06\\ 
 6158.172 & 10.74 & $-$0.995 & B+   & WFD \\
 6158.187 & 10.74 & $-$0.409 & B+   & WFD \\[1mm]
\hline
\end{tabular}
\end{footnotesize}
\end{center}
\end{table}

\addtocounter{table}{-1}
\begin{table}[bh!]
\caption[]{continued.\\[-6mm]}
\setlength{\tabcolsep}{2mm}
\begin{center}
\begin{footnotesize}
\begin{tabular}{lrrlrrr}
\hline\hline
$\lambda$\,({\AA}) & $\chi$\,eV) & $\log gf$ & Acc. & Ref. & $\log \varepsilon$ & $\Delta \log \varepsilon$\\
\hline
 6453.602 & 10.74 & $-$1.288 & C+   & WFD  & 8.80 & $-$0.07\\
 6454.444 & 10.74 & $-$1.066 & C+   & WFD \\
 7001.899 & 10.99 & $-$1.489 & B    & WFD  & 8.74 & $-$0.03\\
 7001.922 & 10.99 & $-$1.011 & B    & WFD \\
 7002.173 & 10.99 & $-$2.664 & B    & WFD \\
 7002.196 & 10.99 & $-$1.489 & B    & WFD \\
 7002.230 & 10.99 & $-$0.741 & B    & WFD \\
 7002.250 & 10.99 & $-$1.364 & B    & WFD \\
 7771.944 &  9.15 &    0.354 & A    & FFT  & 8.71 & {\em $-$0.75}\\
 7774.166 &  9.15 &    0.207 & A    & FFT  & 8.75 & {\em $-$0.71}\\
 7775.388 &  9.15 & $-$0.015 & A    & FFT \\        
 8446.247 &  9.52 & $-$0.468 & B    & FFT  & 8.69 & {\em $-$0.83}\\
 8446.359 &  9.52 &    0.231 & B    & FFT \\
 8446.758 &  9.52 &    0.009 & B    & FFT \\[1mm]
\multicolumn{7}{l}{\ion{Na}{i}:\hfill
$\log \varepsilon_\mathrm{LTE}$\,=\,6.30$\pm$0.03\,(3)}\\[0.5mm]
 5688.193 &  2.10 & $-$1.407 & A    & FFTI & 6.28       & {\ldots} \\
 5688.204 &  2.10 & $-$0.453 & A    & FFTI & \\
 5889.951 &  0.00 &    0.108 & AA   & NIST & {\em 6.84} & {\ldots} \\
 5895.924 &  0.00 & $-$0.194 & AA   & NIST & {\em 6.75} & {\ldots} \\
 8183.256 &  2.10 & $-$0.235 & A+   & FFTI & 6.30       & {\ldots} \\
 8194.790 &  2.10 & $-$0.464 & A    & FFTI & 6.33       & {\ldots} \\
 8194.824 &  2.10 &    0.491 & A+   & FFTI & \\[1mm]
\multicolumn{7}{l}{\ion{Mg}{i}:\hfill$\log
\varepsilon_\mathrm{NLTE}$\,=\,7.56$\pm$0.05\,(5),~~~
$\log \varepsilon_\mathrm{LTE}$\,=\,7.57$\pm$0.07\,(5)}\\[0.5mm]
 4702.991 &  4.35 & $-$0.666 & X    & KB   & 7.58 &    0.01 \\
 5172.684 &  2.71 & $-$0.393 & B+   & NIST & 7.61 & $-$0.03 \\
 5183.604 &  2.72 & $-$0.167 & A    & NIST & 7.55 & $-$0.05 \\
 5528.405 &  4.35 & $-$0.620 & X    & KB   & 7.48 &    0.03 \\
 8806.766 &  4.35 & $-$0.134 & A    & NIST & 7.59 &    0.00 \\[1mm]
\multicolumn{7}{l}{\ion{Mg}{ii}:\hfill$\log
\varepsilon_\mathrm{NLTE}$\,=\,7.54$\pm$0.06\,(8),~~~
$\log \varepsilon_\mathrm{LTE}$\,=\,7.52$\pm$0.04\,(8)}\\[0.5mm]
 3848.211 &  8.86 & $-$1.495 & C    & OP   & 7.47 &    0.00\\
 3848.340 &  8.86 & $-$2.398 & D    & OP   & \\
 4390.514 & 10.00 & $-$1.70  & D    & WSM  & 7.50 &    0.04\\
 4390.572 & 10.00 & $-$0.530 & A    & FFTI & \\
 4427.994 & 10.00 & $-$1.208 & A    & FFTI & 7.53 &    0.04\\
 4433.988 & 10.00 & $-$0.907 & A    & FFTI & 7.54 &    0.04\\
 4481.126 &  8.86 &    0.730 & B    & FW   & 7.51 & $-$0.04\\
 4481.150 &  8.86 & $-$0.570 & B    & FW   & \\
 4481.325 &  8.86 &    0.575 & B    & FW   & \\
 6545.968 & 11.63 &    0.41  & C    & CA   & 7.65 &    0.11\\
 7877.054 & 10.00 &    0.391 & A+   & FFTI & 7.54 & $-$0.01\\
 7896.040 & 10.00 & $-$0.308 & A    & FFTI & 7.58 &    0.01\\
 7896.366 & 10.00 &    0.647 & A+   & FFTI & \\[1mm]
\multicolumn{7}{l}{\ion{Al}{i}:\hfill
$\log \varepsilon_\mathrm{LTE}$\,=\,6.19$\pm$0.07\,(2)}\\[0.5mm]
 3944.006 &  0.00 & $-$0.64  & C+   & WSM  & 6.24 & {\ldots} \\
 3961.520 &  0.01 & $-$0.337 & C+   & WSM  & 6.14 & {\ldots} \\[1mm]
\multicolumn{7}{l}{\ion{Si}{ii}:\hfill$\log
\varepsilon_\mathrm{NLTE}$\,=\,7.57$\pm$0.04\,(7),~~~
$\log \varepsilon_\mathrm{LTE}$\,=\,7.58$\pm$0.05\,(5)}\\[0.5mm]
 3853.665 &  6.86 & $-$1.397 & C    & FFTI & 7.52 & $-$0.03\\
 3862.595 &  6.86 & $-$0.757 & C+   & NIST & 7.59 & $-$0.06\\
 5055.984 & 10.07 &    0.42  & D+   & WSM  & 7.62 &    0.01\\
 5056.317 & 10.07 & $-$0.53  & E    & WSM  & \\
 5957.559 & 10.07 & $-$0.36  & D    & WSM  & 7.59 &    0.02\\
 5978.930 & 10.07 & $-$0.06  & D    & WSM  & 7.56 &    0.03\\
 6347.109 &  8.12 &    0.176 & C+   & FFTI & 7.62 & {\em $-$0.25}\\
 6371.371 &  8.12 & $-$0.126 & C+   & FFTI & 7.53 & {\em $-$0.23}\\[1mm]
\multicolumn{7}{l}{\ion{S}{i}:\hfill
$\log \varepsilon_\mathrm{LTE}$\,=\,7.22$\pm$0.14\,(4)}\\[0.5mm]
 6052.66  &  7.87 & $-$0.393 & X    & FW   & 7.12 & {\ldots} \\   
 6748.58  &  7.87 & $-$0.803 & X    & FW   & 7.21 & {\ldots} \\
 6748.79  &  7.87 & $-$0.596 & X    & FW   & \\
 6756.75  &  7.87 & $-$1.745 & X    & FW   & 7.13 & {\ldots} \\
 6756.96  &  7.87 & $-$0.905 & X    & FW   & \\
 6757.15  &  7.87 & $-$0.307 & X    & FW   & \\
 8693.16  &  7.87 & $-$1.377 & X    & FW   & 7.42 & {\ldots} \\
 8693.98  &  7.87 & $-$0.521 & X    & FW   & \\
 8694.71  &  7.87 &    0.050 & X    & FW   & \\
\hline
\end{tabular}
\end{footnotesize}
\end{center}
\end{table}

\addtocounter{table}{-1}
\begin{table}[bh!]
\caption[]{continued.\\[-6mm]}
\setlength{\tabcolsep}{2mm}
\begin{center}
\begin{footnotesize}
\begin{tabular}{lrrlrrr}
\hline\hline
$\lambda$\,({\AA}) & $\chi$\,eV) & $\log gf$ & Acc. & Ref. & $\log \varepsilon$ & $\Delta \log \varepsilon$\\
\hline
\multicolumn{7}{l}{\ion{Ca}{i}:\hfill
$\log \varepsilon_\mathrm{LTE}$\,=\,6.20$\pm$0.12\,(11)}\\[0.5mm]
 4226.728 &  0.00 &    0.244 & B+   & NIST & 6.48 & {\ldots} \\
 4318.652 &  1.90 & $-$0.21  & C+   & NIST & 6.23 & {\ldots} \\
 4425.437 &  1.88 & $-$0.358 & C    & FW   & 6.25 & {\ldots} \\
 4434.957 &  1.88 & $-$0.01  & C    & FW   & 6.15 & {\ldots} \\
 4581.395 &  2.52 & $-$0.337 & C    & NIST & 6.10 & {\ldots} \\
 4581.467 &  2.52 & $-$1.26  & D    & NIST & \\
 4585.865 &  2.53 & $-$0.187 & C    & NIST & 6.00 & {\ldots} \\
 4585.964 &  2.53 & $-$1.26  & D    & NIST & \\
 4586.036 &  2.53 & $-$2.81  & E    & NIST & \\
 5857.451 &  2.93 &    0.23  & D    & NIST & 6.27 & {\ldots} \\
 6122.217 &  1.89 & $-$0.315 & C    & FW   & 6.18 & {\ldots} \\
 6162.173 &  1.90 & $-$0.089 & C    & FW   & 6.23 & {\ldots} \\
 6439.075 &  2.53 &    0.47  & D    & NIST & 6.19 & {\ldots} \\
 6462.567 &  2.52 &    0.31  & D    & NIST & 6.15 & {\ldots} \\[1mm]
\multicolumn{7}{l}{\ion{Ca}{ii}:\hfill
$\log \varepsilon_\mathrm{LTE}$\,=\,6.27$\pm$0.17\,(9)}\\[0.5mm]
 4799.973 &  8.44 & $-$0.419 & C    & OP   & 6.26 & {\ldots} \\
 5019.971 &  7.51 & $-$0.28  & D    & NIST & 6.28 & {\ldots} \\
 5307.224 &  7.51 & $-$0.90  & D    & NIST & 6.36 & {\ldots} \\
 8248.80  &  7.51 &    0.57  & C    & NIST & 5.95 & {\ldots} \\
 8254.72  &  7.51 & $-$0.39  & C    & NIST & 6.04 & {\ldots} \\
 8498.02  &  1.69 & $-$1.47  & C    & OP   & 6.43 & {\ldots} \\
 8662.14  &  1.69 & $-$0.77  & C    & OP   & 6.39 & {\ldots} \\
 8912.07  &  7.05 &    0.575 & X    & KB   & 6.41 & {\ldots} \\
 8927.36  &  7.05 &    0.750 & X    & KB   & 6.34 & {\ldots} \\[1mm]
\multicolumn{7}{l}{\ion{Sc}{ii}:\hfill
$\log \varepsilon_\mathrm{LTE}$\,=\,3.20$\pm$0.10\,(5)}\\[0.5mm]
 4246.822 &  0.32 &    0.28  & X    & KB   & 3.24 & ... \\
 4324.996 &  0.60 & $-$0.44  & B    & MFW  & 3.20 & {\ldots} \\ 
 5239.813 &  1.46 & $-$0.77  & B    & MFW  & 3.17 & {\ldots} \\
 5526.790 &  1.77 &    0.06  & X    & KB   & 3.06 & {\ldots} \\
 5667.149 &  1.50 & $-$1.19  & B    & MFW  & 3.33 & {\ldots} \\[1mm]
\multicolumn{7}{l}{\ion{Ti}{ii}:\hfill$\log
\varepsilon_\mathrm{NLTE}$\,=\,5.51$\pm$0.17\,(29),~~~
$\log \varepsilon_\mathrm{LTE}$\,=\,5.55$\pm$0.21\,(29)}\\[0.5mm]
 3900.539 &  1.13 & $-$0.20  & B    & P    & 5.63 & $-$0.17\\
 4028.343 &  1.89 & $-$1.00  & D    & MFW  & 5.39 &    0.00\\              
 4163.644 &  2.59 & $-$0.13  & B    & P    & 5.43 & $-$0.18\\              
 4171.904 &  2.60 & $-$0.29  & B    & P    & 5.60 & $-$0.08\\              
 4287.872 &  1.08 & $-$2.02  & D$-$ & MFW  & 5.54 &    0.02\\              
 4294.099 &  1.08 & $-$1.11  & D$-$ & MFW  & 5.86 & $-$0.10\\              
 4300.042 &  1.18 & $-$0.44  & B    & P    & 5.67 & $-$0.07\\              
 4301.922 &  1.16 & $-$1.15  & B    & P    & 5.43 & $-$0.03\\              
 4312.856 &  1.18 & $-$1.10  & B    & P    & 5.51 & $-$0.05\\              
 4395.033 &  1.08 & $-$0.66  & D$-$ & MFW  & 5.80 & $-$0.09\\              
 4399.772 &  1.24 & $-$1.27  & D$-$ & MFW  & 5.39 & $-$0.01\\              
 4443.794 &  1.08 & $-$0.70  & D$-$ & MFW  & 5.51 & $-$0.12\\              
 4450.482 &  1.08 & $-$1.52  & B    & P    & 5.38 &    0.00\\              
 4468.507 &  1.13 & $-$0.60  & D$-$ & MFW  & 5.57 & $-$0.11\\              
 4488.331 &  3.11 & $-$0.82  & D$-$ & MFW  & 5.52 &    0.00\\              
 4501.273 &  1.12 & $-$0.75  & D$-$ & MFW  & 5.61 & $-$0.09\\              
 4529.474 &  1.57 & $-$2.03  & D$-$ & MFW  & 5.67 &    0.02\\              
 4563.761 &  1.22 & $-$0.96  & D$-$ & MFW  & 5.76 &    0.03\\              
 4571.971 &  1.57 & $-$0.32  & B    & P    & 5.81 & $-$0.12\\              
 4589.958 &  1.24 & $-$1.79  & D$-$ & MFW  & 5.47 &    0.00\\              
 4779.985 &  2.05 & $-$1.37  & D$-$ & MFW  & 5.33 &    0.00\\              
 4798.532 &  1.08 & $-$2.68  & C    & P    & 5.35 &    0.02\\              
 4805.085 &  2.06 & $-$1.10  & D$-$ & MFW  & 5.43 & $-$0.03\\              
 4911.193 &  3.12 & $-$0.61  & B    & P    & 5.20 &    0.02\\              
 5010.211 &  3.09 & $-$1.29  & C    & P    & 5.27 &    0.01\\              
 5129.152 &  1.89 & $-$1.39  & D$-$ & MFW  & 5.45 &    0.01\\              
 5185.913 &  1.89 & $-$1.35  & D    & MFW  & 5.24 &    0.00\\              
 5188.680 &  1.58 & $-$1.21  & D$-$ & MFW  & 5.55 & $-$0.06\\              
 5336.771 &  1.58 & $-$1.70  & D$-$ & MFW  & 5.31 &    0.01\\[1mm]
\multicolumn{7}{l}{\ion{V}{ii}:\hfill
$\log \varepsilon_\mathrm{LTE}$\,=\,4.00$\pm$0.17\,(2)}\\[0.5mm]
 4023.388 &  1.80 & $-$0.52  & X    & KB   & 3.88 & {\ldots} \\
 4183.440 &  2.05 & $-$0.95  & X    & KB   & 4.12 & {\ldots} \\
\hline
\end{tabular}
\end{footnotesize}
\end{center}
\end{table}

\addtocounter{table}{-1}
\begin{table}[bh!]
\caption[]{continued.\\[-6mm]}
\setlength{\tabcolsep}{2mm}
\begin{center}
\begin{footnotesize}
\begin{tabular}{lrrlrrr}
\hline\hline
$\lambda$\,({\AA}) & $\chi$\,eV) & $\log gf$ & Acc. & Ref. & $\log \varepsilon$ & $\Delta \log \varepsilon$\\
\hline
\multicolumn{7}{l}{\ion{Cr}{i}:\hfill
$\log \varepsilon_\mathrm{LTE}$\,=\,5.80$\pm$0.01\,(3)}\\[0.5mm]
 4254.331 &  0.00 & $-$0.11  & B    & MFW  & 5.81 & {\ldots} \\
 4274.806 &  0.00 & $-$0.23  & B    & MFW  & 5.79 & {\ldots} \\
 5204.505 &  0.94 & $-$0.20  & B    & MFW  & 5.80 & {\ldots} \\[1mm]
\multicolumn{7}{l}{\ion{Cr}{ii}:\hfill
$\log \varepsilon_\mathrm{LTE}$\,=\,5.95$\pm$0.11\,(27)}\\[0.5mm]
 4037.972 &  6.49 & $-$0.557 & X    & KB   & 5.75 & {\ldots} \\
 4051.930 &  3.09 & $-$2.192 & X    & KB   & 5.77 & {\ldots} \\
 4242.364 &  3.87 & $-$1.331 & X    & KB   & 6.08 & {\ldots} \\
 4252.632 &  3.84 & $-$2.018 & X    & KB   & 6.05 & {\ldots} \\
 4261.913 &  3.87 & $-$1.531 & X    & KB   & 5.93 & {\ldots} \\
 4275.567 &  3.86 & $-$1.709 & X    & KB   & 6.01 & {\ldots} \\
 4284.188 &  3.86 & $-$1.864 & X    & KB   & 6.02 & {\ldots} \\
 4588.199 &  4.07 & $-$0.63  & D    & MFW  & 6.05 & {\ldots} \\
 4592.049 &  4.07 & $-$1.22  & D    & MFW  & 5.85 & {\ldots} \\
 4616.629 &  4.07 & $-$1.29  & D    & MFW  & 5.82 & {\ldots} \\
 4634.070 &  4.07 & $-$1.24  & D    & MFW  & 6.12 & {\ldots} \\
 4812.337 &  3.86 & $-$1.99  & X    & KP   & 5.95 & {\ldots} \\
 4824.127 &  3.87 & $-$0.96  & X    & KP   & 6.00 & {\ldots} \\
 4836.229 &  3.86 & $-$1.94  & X    & KP   & 6.01 & {\ldots} \\
 4848.235 &  3.86 & $-$1.14  & X    & KP   & 6.03 & {\ldots} \\
 4876.399 &  3.86 & $-$1.46  & D    & MFW  & 6.00 & {\ldots} \\
 5237.329 &  4.06 & $-$1.16  & D    & MFW  & 5.94 & {\ldots} \\
 5246.768 &  3.71 & $-$2.45  & D    & MFW  & 5.97 & {\ldots} \\
 5249.437 &  3.76 & $-$2.426 & X    & KB   & 5.85 & {\ldots} \\
 5274.964 &  4.05 & $-$1.29  & X    & KB   & 5.94 & {\ldots} \\
 5279.876 &  4.07 & $-$2.10  & D    & MFW  & 5.88 & {\ldots} \\
 5308.408 &  4.07 & $-$1.81  & D    & MFW  & 5.81 & {\ldots} \\
 5310.686 &  4.07 & $-$2.28  & D    & MFW  & 5.99 & {\ldots} \\
 5313.563 &  4.07 & $-$1.65  & D    & MFW  & 6.02 & {\ldots} \\
 5334.869 &  4.05 & $-$1.56  & X    & KB   & 5.85 & {\ldots} \\
 5508.606 &  4.16 & $-$2.11  & D    & MFW  & 5.90 & {\ldots} \\
 5510.702 &  3.83 & $-$2.452 & X    & KB   & 6.01 & {\ldots} \\[1mm]
\multicolumn{7}{l}{\ion{Mn}{i}:\hfill
$\log \varepsilon_\mathrm{LTE}$\,=\,5.48$\pm$0.15\,(4)}\\[0.5mm]
 4034.483 &  0.00 & $-$0.811 & C+   & MFW  & 5.34 & {\ldots} \\
 4041.355 &  2.11 &    0.285 & C+   & MFW  & 5.50 & {\ldots} \\
 4823.524 &  2.32 &    0.144 & C+   & MFW  & 5.48 & {\ldots} \\
 6021.819 &  3.08 &    0.034 & C+   & MFW  & 5.69 & {\ldots} \\[1mm]
\multicolumn{7}{l}{\ion{Mn}{ii}:\hfill
$\log \varepsilon_\mathrm{LTE}$\,=\,5.88$\pm$0.10\,(2)}\\[0.5mm]
 4206.367 &  5.40 & $-$1.566 & X    & KB   & 5.81 & {\ldots} \\
 7415.810 &  3.71 & $-$2.202 & X    & KB   & 5.95 & {\ldots} \\[1mm]
\multicolumn{7}{l}{\ion{Fe}{i}:\hfill
$\log \varepsilon_\mathrm{LTE}$\,=\,7.65$\pm$0.11\,(45)}\\[0.5mm]
 3922.912 &  0.05 & $-$1.651 & B+   & FMW  & 7.68 & {\ldots} \\
 3927.920 &  0.11 & $-$1.59  & C    & FMW  & 7.64 & {\ldots} \\
 4009.713 &  2.21 & $-$1.20  & C    & FMW  & 7.53 & {\ldots} \\
 4021.866 &  2.76 & $-$0.66  & C+   & FMW  & 7.47 & {\ldots} \\
 4063.594 &  1.56 &    0.07  & C+   & FMW  & 7.84 & {\ldots} \\
 4071.738 &  1.61 & $-$0.022 & B+   & FMW  & 7.64 & {\ldots} \\
 4118.545 &  3.57 &    0.28  & C    & FMW  & 7.56 & {\ldots} \\
 4143.415 &  3.05 & $-$0.470 & X    & KB   & 7.74 & {\ldots} \\
 4143.868 &  1.56 & $-$0.45  & C+   & FMW  & \\
 4187.039 &  2.45 & $-$0.548 & B+   & FMW  & 7.51 & {\ldots} \\
 4187.795 &  2.43 & $-$0.554 & B+   & FMW  & 7.69 & {\ldots} \\
 4219.360 &  3.56 &    0.12  & C+   & FMW  & 7.50 & {\ldots} \\
 4222.213 &  2.44 & $-$0.967 & B+   & FMW  & 7.55 & {\ldots} \\
 4235.936 &  2.41 & $-$0.341 & B+   & FMW  & 7.61 & {\ldots} \\
 4250.119 &  2.47 & $-$0.405 & B+   & FMW  & 7.55 & {\ldots} \\
 4250.787 &  1.58 & $-$0.71  & D$-$ & FMW  & 7.53 & {\ldots} \\
 4278.230 &  3.37 & $-$1.74  & C    & FMW  & 7.54 & {\ldots} \\
 4325.762 &  1.61 & $-$0.01  & C+   & FMW  & 7.79 & {\ldots} \\
 4447.717 &  2.22 & $-$1.34  & B+   & FMW  & 7.54 & {\ldots} \\
 4459.117 &  2.18 & $-$1.279 & B+   & FMW  & 7.81 & {\ldots} \\
 4484.220 &  3.59 & $-$0.72  & D    & FMW  & 7.64 & {\ldots} \\
 4494.563 &  2.20 & $-$1.136 & B+   & FMW  & 7.58 & {\ldots} \\
 4903.310 &  2.88 & $-$1.08  & C    & FMW  & 7.64 & {\ldots} \\
\hline
\end{tabular}
\end{footnotesize}
\end{center}
\end{table}

\addtocounter{table}{-1}
\begin{table}[bh!]
\caption[]{continued.\\[-6mm]}
\setlength{\tabcolsep}{2mm}
\begin{center}
\begin{footnotesize}
\begin{tabular}{lrrlrrr}
\hline\hline
$\lambda$\,({\AA}) & $\chi$\,eV) & $\log gf$ & Acc. & Ref. & $\log \varepsilon$ & $\Delta \log \varepsilon$\\
\hline
 4918.994 &  2.87 & $-$0.37  & C+   & FMW  & 7.56 & {\ldots} \\
 4920.502 &  2.83 &    0.06  & C+   & FMW  & 7.76 & {\ldots} \\
 5068.766 &  2.94 & $-$1.23  & C    & FMW  & 7.81 & {\ldots} \\
 5133.688 &  4.18 &    0.14  & D    & FMW  & 7.79 & {\ldots} \\
 5266.555 &  3.00 & $-$0.49  & C+   & FMW  & 7.83 & {\ldots} \\
 5281.790 &  3.04 & $-$1.02  & C    & FMW  & 7.72 & {\ldots} \\
 5324.179 &  3.21 & $-$0.24  & C+   & FMW  & 7.65 & {\ldots} \\
 5328.039 &  0.91 & $-$1.466 & B+   & FMW  & 7.64 & {\ldots} \\
 5328.531 &  1.56 & $-$1.65  & X    & KB   & \\
 5364.871 &  4.45 &    0.22  & D    & FMW  & 7.53 & {\ldots} \\
 5367.466 &  4.42 &    0.35  & C+   & FMW  & 7.60 & {\ldots} \\
 5369.961 &  4.37 &    0.35  & C+   & FMW  & 7.75 & {\ldots} \\
 5383.369 &  4.31 &    0.50  & C+   & FMW  & 7.72 & {\ldots} \\
 5397.128 &  0.91 & $-$1.993 & B+   & FMW  & 7.60 & {\ldots} \\
 5405.775 &  0.99 & $-$1.844 & B+   & FMW  & 7.54 & {\ldots} \\
 5410.910 &  4.47 &    0.28  & C+   & FMW  & 7.66 & {\ldots} \\
 5424.068 &  4.32 &    0.52  & D    & FMW  & 7.77 & {\ldots} \\
 5445.042 &  4.39 & $-$0.02  & D    & FMW  & 7.64 & {\ldots} \\
 5569.618 &  3.42 & $-$0.54  & C+   & FMW  & 7.56 & {\ldots} \\
 5572.842 &  3.40 & $-$0.31  & C+   & FMW  & 7.50 & {\ldots} \\
 5586.756 &  3.37 & $-$0.21  & C+   & FMW  & 7.66 & {\ldots} \\
 5615.644 &  3.33 & $-$0.14  & C+   & FMW  & 7.64 & {\ldots} \\
 6024.058 &  4.55 & $-$0.12  & D$-$ & FMW  & 7.81 & {\ldots} \\
 6400.000 &  3.60 & $-$0.52  & D$-$ & FMW  & 7.78 & {\ldots} \\[1mm]
\multicolumn{7}{l}{\ion{Fe}{ii}:\hfill$\log
\varepsilon_\mathrm{NLTE}$\,=\,7.70$\pm$0.10\,(34),~~~
$\log \varepsilon_\mathrm{LTE}$\,=\,7.68$\pm$0.10\,(34)}\\[0.5mm]
 4122.668 &  2.58 & $-$3.38  & D    & FMW  & 7.64 & 0.02\\
 4173.461 &  2.58 & $-$2.18  & C    & FWW  & 7.79 & 0.02\\
 4273.326 &  2.70 & $-$3.34  & D    & FMW  & 7.63 & 0.02\\
 4296.572 &  2.70 & $-$3.01  & D    & FMW  & 7.71 & 0.02\\
 4303.176 &  2.70 & $-$2.49  & C    & FMW  & 7.60 & 0.03\\
 4416.830 &  2.78 & $-$2.60  & D    & FMW  & 7.62 & 0.03\\
 4472.929 &  2.84 & $-$3.43  & D    & FMW  & 7.54 & 0.02\\
 4489.183 &  2.83 & $-$2.97  & D    & FMW  & 7.71 & 0.02\\
 4508.288 &  2.86 & $-$2.31  & D    & KB   & 7.80 & 0.02\\
 4515.339 &  2.84 & $-$2.48  & D    & FMW  & 7.74 & 0.02\\
 4520.224 &  2.81 & $-$2.60  & D    & FMW  & 7.78 & 0.02\\
 4522.634 &  2.84 & $-$2.11  & C    & KB   & 7.85 & 0.02\\
 4555.893 &  2.83 & $-$2.32  & D    & KB   & 7.55 & 0.03\\
 4576.340 &  2.84 & $-$3.04  & D    & FMW  & 7.68 & 0.03\\
 4620.521 &  2.83 & $-$3.28  & D    & FMW  & 7.53 & 0.02\\
 4629.339 &  2.81 & $-$2.37  & D    & FMW  & 7.85 & 0.01\\
 4635.316 &  5.96 & $-$1.65  & D$-$ & FMW  & 7.68 & 0.01\\
 4666.758 &  2.83 & $-$3.33  & D    & FMW  & 7.65 & 0.02\\
 4731.453 &  2.89 & $-$3.36  & D    & FMW  & 7.77 & 0.02\\
 4993.358 &  2.81 & $-$3.65  & E    & FMW  & 7.56 & $-$0.01\\
 5169.033 &  2.89 & $-$0.87  & C    & FMW  & 7.61 & $-$0.05\\
 5197.577 &  3.23 & $-$2.23  & X    & KB   & 7.72 & $-$0.02\\
 5276.002 &  3.20 & $-$1.94  & C    & FMW  & 7.83 & 0.00\\
 5325.553 &  3.22 & $-$3.22  & X    & KB   & 7.67 & 0.02\\
 5425.257 &  3.20 & $-$3.36  & D    & FMW  & 7.73 & 0.02\\
 5427.826 &  6.72 & $-$1.66  & X    & KB   & 7.78 & 0.02\\
 5534.847 &  3.24 & $-$2.93  & D    & FMW  & 7.83 & 0.02\\
 6084.111 &  3.20 & $-$3.98  & D    & FMW  & 7.82 & 0.01\\
 6147.741 &  3.89 & $-$2.721 & X    & KB   & 7.68 & 0.02\\
 6149.258 &  3.89 & $-$2.724 & X    & KB   & 7.62 & 0.02\\
 6238.392 &  3.89 & $-$2.630 & X    & KB   & 7.60 & 0.02\\
 6247.557 &  3.89 & $-$2.51  & D    & FMW  & 7.79 & 0.02\\
 6416.919 &  3.89 & $-$2.85  & D    & FMW  & 7.71 & 0.02\\
 6432.680 &  2.89 & $-$3.74  & D    & FMW  & 7.84 & 0.02\\[1mm]
\multicolumn{7}{l}{\ion{Co}{ii}:\hfill
$\log \varepsilon_\mathrm{LTE}$\,=5.32$\pm$\,0.03\,(2)}\\[0.5mm]
 4160.673 &  3.41 & $-$1.828 & X   & KB   & 5.30 & {\ldots} \\
 4660.656 &  3.36 & $-$2.205 & X   & KB   & 5.34 & {\ldots} \\
\hline
\end{tabular}
\end{footnotesize}
\end{center}
\end{table}

\addtocounter{table}{-1}
\begin{table}[bh!]
\caption[]{continued.\\[-6mm]}
\setlength{\tabcolsep}{2mm}
\begin{center}
\begin{footnotesize}
\begin{tabular}{lrrlrrr}
\hline\hline
$\lambda$\,({\AA}) & $\chi$\,eV) & $\log gf$ & Acc. & Ref. & $\log \varepsilon$ & $\Delta \log \varepsilon$\\
\hline
\multicolumn{7}{l}{\ion{Ni}{i}:\hfill
$\log \varepsilon_\mathrm{LTE}$\,=\,6.51$\pm$0.10\,(6)}\\[0.5mm]
 3858.297 &  0.42 & $-$0.97  & C   & FMW  & 6.36 & {\ldots} \\
 4604.987 &  3.48 & $-$0.29  & D$-$& FMW  & 6.54 & {\ldots} \\
 4714.417 &  3.37 &    0.23  & D   & FMW  & 6.42 & {\ldots} \\
 4786.535 &  3.42 & $-$0.17  & D   & FMW  & 6.64 & {\ldots} \\ 
 4980.173 &  3.61 & $-$0.11  & D   & FMW  & 6.54 & {\ldots} \\
 5115.392 &  3.83 & $-$0.11  & D$-$& FMW  & 6.57 & {\ldots} \\[1mm]
\multicolumn{7}{l}{\ion{Ni}{ii}:\hfill
$\log \varepsilon_\mathrm{LTE}$\,=6.63\,(1)}\\[0.5mm]
 4015.474 &  4.03 & $-$2.42  & X   & KB   & 6.63 & {\ldots} \\[1mm]
\multicolumn{7}{l}{\ion{Cu}{i}:\hfill
$\log \varepsilon_\mathrm{LTE}$\,=\,4.47$\pm$0.11\,(2)}\\[0.5mm]
 5153.230 &  3.79 &    0.46  & X   & FW   & 4.54 & {\ldots} \\
 5218.197 &  3.82 &    0.44  & X   & FW   & 4.39 & {\ldots} \\[1mm]
\multicolumn{7}{l}{\ion{Zn}{i}:\hfill
$\log \varepsilon_\mathrm{LTE}$\,=\,5.16$\pm$0.06\,(2)}\\[0.5mm]
 4722.153 &  4.03 & $-$0.37  & X   & KSZ  & 5.20 & {\ldots} \\
 4810.528 &  4.06 & $-$0.15  & X   & KSZ  & 5.12 & {\ldots} \\[1mm]
\multicolumn{7}{l}{\ion{Sr}{ii}:\hfill
$\log \varepsilon_\mathrm{LTE}$\,=\,3.18$\pm$0.03\,(2)}\\[0.5mm]
 4077.709 &  0.00 &    0.15  & X   & FW   & 3.16 & {\ldots} \\
 4215.519 &  0.00 & $-$0.17  & X   & FW   & 3.20 & {\ldots} \\[1mm]
\multicolumn{7}{l}{\ion{Y}{ii}:\hfill
$\log \varepsilon_\mathrm{LTE}$\,=\,2.14$\pm$0.23\,(3)}\\[0.5mm]
 3774.339 &  0.13 &    0.22  & X   & H    & 2.40 & {\ldots} \\
 3788.697 &  0.10 & $-$0.06  & X   & H    & 2.05 & {\ldots} \\
 3950.356 &  0.10 & $-$0.49  & X   & H    & 1.98 & {\ldots} \\[1mm]
\multicolumn{7}{l}{\ion{Zr}{ii}:\hfill
$\log \varepsilon_\mathrm{LTE}$\,=\,3.05.$\pm$0.08\,(3)}\\[0.5mm]
 4050.329 &  0.71 & $-$0.99  & X   & B    & 2.98 & {\ldots} \\
 4149.202 &  0.80 & $-$0.03  & X   & B    & 3.13 & {\ldots} \\
 4496.974 &  0.71 & $-$0.81  & X   & B    & 3.05 & {\ldots} \\[1mm]
\multicolumn{7}{l}{\ion{Ba}{ii}:\hfill
$\log \varepsilon_\mathrm{LTE}$\,=\,3.00$\pm$0.18\,(5)}\\[0.5mm]
 4554.029 &  0.00 &    0.14  & X   & D    & 3.21 & {\ldots} \\
 4934.076 &  0.00 & $-$0.16  & X   & D    & 3.15 & {\ldots} \\
 5853.668 &  0.60 & $-$0.51  & X   & D    & 2.77 & {\ldots} \\
 6141.713 &  0.70 &    0.19  & X   & D    & 2.89 & {\ldots} \\
 6496.897 &  0.60 & $-$0.01  & X   & D    & 3.00 & {\ldots} \\  
\hline
\end{tabular}
\tablefoot{
Accuracy indicators -- uncertainties within
AAA: 0.3\%;
AA:  1\%;
A:   3\%;
B:  10\%;
C:  25\%;
D:  50\%;
E:  larger than 50\%;
X:  unknown.}
\tablebib{
B:~Biemont, E., et al. 1981, \apj, 248, 867; 
CA:~Coulomb approximation, Bates, D., \& Damgaard, A., 1949, Phil. Trans. Roy. Soc., 242A, 101; 
D:~Davidson, M. D., et al. 1992, \aap, 255, 457; 
FFT:~Froese Fischer, C., \& Tachiev, G. 2004, At. Data Nucl. Data Tables, 87, 1; 
FFTI:~Froese Fischer, C., et al. 2006, At. Data Nucl. Data Tables, 92, 607; 
FMW:~Fuhr, J. R., et al. 1988, J. Phys. \& Chem. Ref. Data, Vol. 17, Suppl. 4; 
FW:~Fuhr, J. R., \& Wiese, W. L., 1998, in CRC Handbook of Chemistry and Physics, 79th edn., ed. D. R. Lide (Boca Raton: CRC Press); 
H:~Hannaford, P., et al. 1982, \apj, 261, 736; 
KB:~Kurucz, R. L., \& Bell, B. 1995, CD-ROM No. 23 (Cambridge, Mass: SAO); 
KSZ:~Kerkhoff, H., et al. 1980, Z. Phys. A, 298, 249; 
LP:~Luo, D., \& Pradhan, A. K. 1989, J. Phys. B, 22, 3377; 
MFW:~Martin, G. A., et al. 1988, J. Phys. Chem. Ref. Data, 17, Suppl. 3; 
NIST:~Kramida, A., et al. 2016, NIST Atomic Spectra Database (v5.4), http://physics.nist.gov/asd (NIST, Gaithersburg, MD); 
OP:~Opacity Project, available electronically via TOPBASE (http://cdsweb.u-strasbg.fr/topbase/topbase.html);
P:~Pickering, J. C., et al. 2001, \apjs, 132, 403; Pickering, J. C., et al. 2002, \apjs, 138, 247;
WF:~Wiese, W. L., \& Fuhr, J. R. 2009, J. Phys. Chem. Ref. Data, 38, 565; 
WFD:~Wiese, W. L., et al. 1996, J. Phys. \& Chem. Ref. Data, Mon., 7; 
WSM:~Wiese, W. L., et al. 1969 Nat. Stand. Ref. Data Ser., Nat. Bur.
Stand. (U.S.), NSRDS-NBS 22, Vol. II.} 
\end{footnotesize}
\end{center}
\end{table}

\end{appendix}
\end{document}